\documentclass[12pt]{article}

\usepackage{graphicx}
\usepackage[utf8]{inputenc}

\usepackage[english]{babel}

\usepackage{a4}
\usepackage{amsmath,amssymb,amsfonts,amsthm} 
\usepackage{mathtools}
\usepackage{cite}
\usepackage{epsfig}

\numberwithin{equation}{section}

\newcommand{\bA}{\textbf{A}: \ } 
\newcommand{\bB}{\textbf{B}: \ }

\def\ie{{\it i.e.\ }}

\newcommand{\be}{\begin{equation}}
\newcommand{\ee}{\end{equation}} 

\newcommand{\mn}{\medskip \noindent}

\begin{document}

\title{\Large  
  Proper condensates versus the Onsager-Penrose criterion: \\  
A friendly debate on Bose-Einstein condensation}
\author{Detlev Buchholz \\[2mm]
\small Mathematisches Institut, Universit\"at G\"ottingen, \\
\small Bunsenstr.\ 3-5, 37073 G\"ottingen, Germany \\ 
\small detlev.buchholz@mathematik.uni-goettingen.de \\[5pt]
}
\date{}

\maketitle

\noindent \textbf{Abstract.}
This article contrasts the concepts of a proper condensate and
the Onsager-Penrose criterion for Bose-Einstein condensation
in the form of a debate between two proponents of the respective concepts.
The prologue briefly introduces the two criteria.
The epilogue contains remarks on 
properties of the underlying resolvent algebra 
that are physically relevant in this context.

\section{Prologue}
\label{sec1}
\setcounter{equation}{0}

This is a transcript of a debate between two older theoretical 
physicists, both of whom have had university careers.
The topic was the phenomenon of Bose-Einstein condensation. 
Since they wish to remain
anonymous, I will refer to them as Alice and Bob.

\medskip 
Alice has been working in the field of condensed matter physics for
about three decades, focusing in particular on the phenomenon of
Bose-Einstein condensation.  She regards the Onsager-Penrose (OP)
criterion \cite{OnPe} as a fundamental
basis of her work.

\medskip
This criterion involves the number operator $N(f)$, which counts how many
particles (bosons) with wave function $f$ appear in a quantum state.   
When dealing with a sequence of $n$-particle states,
such as canonical ensembles at a
given temperature, one considers the average value 
$\langle N(f) \rangle_n$ in these 
states, \textit{i.e.}, the average number of particles with 
wave function $f$. The OP-criterion says: If this average increases
with the number $n$ of particles like
$\kappa n$ for some $0 < \kappa \leq 1$, then there exists a 
macroscopically visible condensate in the states in the limit of
large $n$. The largest possible $\kappa$ indicates its fraction.

\medskip
Bob has primarily worked on conceptual problems in different areas of
quantum  physics and quantum field theory. Four years ago he came across
the new concept of a proper condensate (PC) \cite{Bu4} and was intrigued by
it.

\medskip
In this approach one also considers the particle number operators
$N(f)$, but one proceeds to their inverses. To be precise, since $N(f)$ is 
$0$ if there is no particle with wave function $f$ around, one considers
the inverse of $(\mu + N(f))$ for  given number
$\mu > 0$. It is the resolvent of $N(f)$. One then says that a
sequence of $n$-particle states for which 
the expectation values $\langle (\mu + N(f))^{-1} \rangle_n$
approach~$0$ for large $n$ contains a proper condensate.
As with the OP-criterion, it is not easy to determine the exact 
wave function of the particles forming the condensate. In the PC approach one 
proposes to determine it from the space of functions $f$ for which the limit
is \textit{not} $0$. Its orthogonal complement defines the wave
function(s) of the
condensate. 

\section{The debate}
\label{sec2}
\setcounter{equation}{0} 

\mn
\bA You talk about proper condensates. Are there improper condensates?

\mn
\bB If you are thinking about physics, my answer is no. But we are
taking about theoretical physics. And there one must carefully check
whether some concept covers what can be regarded as a condensate. 

\mn
\bA  The OP criterion has been around for 70 years and forms the basis
for a wealth of important results on the properties of condensates.
So why introduce yet another criterion?

\mn 
\bB  The concept of a PC is not a new criterion for identifying
condensates. It is based on the idea that for a true condensate
it should become easy to identify its presence if its density
becomes arbitrarily large.
Its properties, such as the condensate wave functions, can then
be derived from a study of the remaining excitations. The
orthogonal complement of their wave functions determines
the wave function (state) of the condensate.

\mn
\bA  This is obvious, and it is exactly what the OP-criterion says:
Compute the largest eigenvalue $\kappa$ of the one-particle
density matrix, scaled by the inverse of the particle number, and its
wave function. Then you are done.

\mn
\bB Yes. But you could save lots of work if you only wish to
prove that there is such a $\kappa$. Moreover, as you know very well, the
corresponding wave function may not be unique. 

\mn
\bA  And how do you do that?

\mn
\bB  Let me try to sketch the argument. Pick any $f$ which has support
in a region where condensates are studied. One then considers
the scaled resolvent $n (n + N(f))^{-1}$. They are monotonically
increasing with $n$ and bounded by the unit operator. The
criterion for the presence of a condensate in a sequence of
$n$-particle states is that the corresponding expectation values
$\langle n (n + N(f))^{-1} \rangle_n$ are smaller than $1$
in the limit of large $n$.

\mn
\bA And why does this indicate the presence of a condensate?

\mn 
\bB It shows that the OP-criterion for condensation is satisfied. That the 
method works follows from the Cauchy-Schwartz inequality. It implies that
\[
\langle n (n + N(f))^{-1} \rangle_n  \geq
(1 + n^{-1} \langle N(f) \rangle_n )^{-1} \, .
\]
Hence if the left hand side is less than $1$ in the limit,
the right hand side must
be bigger than $(1 + \kappa_f)^{-1}$ for some 
$\kappa_f > 0$. According to the
OP-criterion, this indicates the presence of a condensate. Of course,
$\kappa_f$ is in general smaller than the largest eigenvalue $\kappa$ of the
scaled one-particle density matrix. But since it is positive, you are done. 
You may call $\sigma_f = \kappa_f / \kappa$
the sensitivity of $N(f)$ for the OP-condensate. 

\mn
\bA And what if $\sigma_f$ is $0$?

\mn
\bB Then there is no OP-condensate.

\mn
\bA But I could give you a function $f$ which is orthogonal to the
condensate wave function (or functions, if there are several). 
Then your $\sigma_f$ would be $0$.

\mn
\bB It seems, but you would not manage.
Because you would have to compute the
exact wave function of the condensate, which in general is not easy,
as you know.
Then you could present the result $f$ of your
computations to me. But if 
your $f$ has, by mistake,
a tiny component in the direction of the condensate
wave function, no matter how small, $\sigma_f$ would be different from $0$.

\mn
More seriously, mathematician have developed a measure theory on
Hilbert spaces. According to them the measure of the orthogonal complement
of a closed, proper subspace of a Hilbert space is $0$, regardless of
whether its dimension is finite or infinite. 
Hence, if you choose $f$ at random, the probability that 
$\sigma_f = 0$ is $0$. 

\mn
I am not a computational physicist. But if I were and would be interested
in checking the OP-criterion, I simply would plug a Gaussian into the
resolvents and turn on my computer.

\mn
\bA But why using these resolvents? 

\mn
\bB You could choose other functions. But the resolvents lead to  
conceptual clarity. I did not yet  tell you that the resolvents
$ \mu  (\mu  + N(f))^{-1} $ are almost 
projection operators. To be precise, in any given state and also 
in its excitations, this sequence converges for large
$\mu$ to some projection operator
$P(f)$ which commutes with all other resolvents \cite{Bu6}.
It is a central projection, which is considered a
classical observable. In a single measurement, its value 
is either $0$ or $1$. 

\mn 
Now assume that 
in some state the expectation value of $P(f)$ lies between these
two values. It would imply that the state can 
uniquely be decomposed into a mixture of two states, where in one state $P(f)$
takes the non-fluctuating value $0$, in the other the value $1$.
In the former state the resolvent vanishes, it contains a proper
condensate. In the latter state one finds
a limited number of particles with wave function $f$.
The probability of finding more
than $m$ particles tends to~$0$ for large $m$.

\mn
\bA That is theory. You will not be able to prepare such a state.

\mn
\bB Yes, it is theory. But you would be able to come close to it
in practice if you are dealing with bosons which, in the limit 
of large particle numbers, comply with the OP-criterion for
condensation, say. 
Then you should add more and more particles into a container.
If the force between them is attractive,  
the particles will bind together and form a cluster about some
point. The expectation values of the
scaled resolvents would then be close to~$0$. Alternatively, you could
try to remove all particles from the container and try to prepare a vacuum.
Then the scaled resolvents would give expectation values which are close to $1$.
The first experiment is carried out on weekdays, the second one on
weekends. If one combines the resulting data, they describe a
mixture where, in the limit, $P(f)$ would have
an expectation value close to $2/7$. 

\mn
\bA This is just the standard procedure for preparing mixtures.

\mn
\bB But nature produces such mixtures.

\mn
\bA Yes. But it is a weird idea that she puts an
infinite number of particles into a
fixed container. One has to enlarge it. 

\mn
\bB So you want to proceed to the thermodynamic limit and escape
with the particles to spatial infinity? I refrain from asking you
why that idea is less weird. Of course you know that   
the wave function of the condensate would tend weakly to $0$ in this
limit.

\mn
In order to count the members of the condensate in this limit, one has
to increase the sensitivity of the measuring device.
It is accomplished in theory by 
scaling the number operator $N(f)$
by a factor $V$, the volume of the container. This simply means that,
instead of scaling the resolvents by the particle number $n$, 
you have to scale them by the particle density $n/V$. 
Then you proceed exactly as before. In order to check whether
the condensate is proper, you need to increase now     
the particle density. 

\mn
\bA Which, in turn, is a very unusual idealization.

\mn 
\bB It is for the sake of theoretical clarity. 
Let me summarize: The PC-condition aims at a more specific characterization
of true condensates on the theoretical side.
One checks it in the hypothetical limit of an infinite number of
particles in a finite container or of infinite densities in the
thermodynamic limit. In other words: It is assumed that
the wave function of the condensate does 
not change its shape in the limit. There remains a space of
regular wave functions in its orthogonal complement, and
{vice versa}. One may then say that
such a stable condensate is a proper condensate.
That is the idea underlying the PC-condition. 

\mn
\bA This notion might please you as a theoretician. 
But it is not relevant in physics. Because in actual
experiments nobody counts particles. One analyzes condensates in momentum
space (time of flight imaging). In the theoretical setting one must
consider the off-diagonal elements of the one-particle density
matrices and establish long range correlations between Bose 
fields in the condensate (ODLRO). 
I doubt it that the resolvents of the particle number operator are of
use in this context. 

\mn
\bB Let me remind you that
one considers not only a single resolvent, one uses the algebra
of all sums and finite products of them. And one knows how to recover
from it the fields \cite{Bu2}. However, this step is not needed 
in order to describe the phenomenon of ODLRO. If you want to study
a condensate in a large container, which satisfies the OP-criterion, 
simply choose two wave functions 
$f_1$, $f_2$ which have their supports in distant regions. 
A little algebra shows that the difference between the 
resolvents, attached to the sum and difference 
of these functions, is given by 
\begin{eqnarray*}
&  n (n + N(f_1 - f_2))^{-1} - n  (n + N(f_1 + f_2))^{-1} &  \\   
& = n  (n + N(f_1 - f_2))^{-1} \, n^{-1} \big(N(f_1 + f_2) - N(f_1 - f_2)\big) \,  
  n  (n + N(f_1 + f_2))^{-1} & .
\end{eqnarray*}
But $N(f) = a(f)^*a(f)$, \ie the product of the creation and annihilation
operators of a particle with wave function $f$, which are real
linear. Evaluating this difference in a sequence of $n$-particle states 
satisfying the OP-condition, one obtains in the
limit of large~$n$  
\begin{eqnarray*}
  2 (1 - \kappa_{(f_1 - f_2)})^{-1} \, 
  n^{-1}\langle a^*(f_1) a(f_2) + a^*(f_2) a(f_1)
    \rangle_n (1 - \kappa_{(f_1 + f_2)})^{-1} \, .
\end{eqnarray*}
I have simplified matters and replaced the resolvents by
their mean values in the states 
since they are close to being classical projections
in the limit.
So the off-diagonal elements of the
scaled one-particle density matrix appear  and 
long range correlations
between the fields can be computed by using only resolvents.   

\mn
Let me remark that 
it pops out from this formula that the correlations  become smaller if
one increases the density of the condensate, since the resolvent factors
decrease. As you know, this is also observed in experiments. It is
in accord with the PC-condition, which says 
that quantum correlations would be completely suppressed if one could
proceed to infinite densities. In \cite{Bu5}, conditions were investigated
which imply that the correlations are visible in large bounded regions. 

\mn
\bA So you seem to think that the PC-condition is a conceptual
improvement of the OP-criterion.   
But unless somebody convinces me that I can learn
something new about physics from it, I see no reason to
concern myself with it

\mn
\bB There is the possibility that the OP-criterion is too restrictive and does
not cover all possible manifestations of true condensates.

\mn
\bA You mean that the maximal value of $\langle N_f \rangle_n$ need not
behave like $\kappa n$? There is a model where it behaves like
$\kappa n^{1/2}$. I do not regard that as a condensate. 

\mn
\bB I do not want to argue with you about definitions.
But you know, of course, that particles in a gas and
in a condensate act differently in response to 
external forces.
When appropriate forces are applied, particles can condense on the
walls of a large container, while the majority of the particles
forms a gas in its interior. Since the surface area  
divided by the volume approaches $0$ for large containers, I
guess that the OP-criterion would indicate that there is no
condensate. The PC-condition would. 

\mn
\bA Do you have a model?

\mn
\bB Such a model was presented in a recent paper
\cite{Bu6}. There,  
non-interacting bosons are considered.
Under suitable conditions, they 
form condensates which are not covered by the OP-criterion, but
they satisfy the PC-condition. 
The particles are exposed to an external force such that 
the wave function of the ground state is constant in a sphere of given radius
and decays rapidly at infinity;
furthermore, there is a countable number of
excited states. The model describes a container with soft
boundaries, which has already been studied in \cite{BuYn}.

\mn 
\bA I have seen the latter paper.

\mn 
\bB But now one considers different sequences of states.
One constructs $n$-particle states, in which $m(n)$ particles occupy
the ground state and each of the remaining
$(n - m(n))$ particles occupies a different one of the orthonormalized
excited states. Thus, whenever $f$ is orthogonal to the ground state,
the expectation value of $N(f)$ is less than $1$ in these states.  
But if $f$ has a component in its direction, it diverges
like $n^{1/2}$.

\mn
\bA And why does that define a condensate?

\mn
\bB It satisfies the $PC$-condition; because there is a 
space of wave functions for which the (unscaled) 
resolvents do not vanish
in the limit. The infinitely populated ground state
is determined by the orthogonal complement. 
Thus, by definition, it is a proper condensate.  
The OP-criterion (and you) say that it can not be interpreted as
a condensate. 

\mn
\bA And why should I interpret it as a true condensate?

\mn 
\bB  To see this, one has to proceed to 
the idealization that $n$ approaches infinity.
As I tried to explain, the proper condensate
becomes a classical system; all quantum observables which are
sensitive to it have a sharp,
non-fluctuating value, which is set equal to 0.
(Let me mention as an aside that the resolvents would now have to
be scaled by $n^{1/2}$ in order to approximate in the sequence
of states the
central projections $P(f)$, which distinguish the phases.) 
So the condensed system is
just there, like the walls of a laboratory.

\mn 
Now imagine there is an observer
inside the sphere whose observables are the resolvents with
wave functions in the space generated by all excited states. 
Due to the specific choice of the ground state, 
she can move around her observables inside the
sphere and examine the excitations. And she would not see
the particles in the
ground state. If, however, she were to try to cross the
boundary of the sphere and move any of her resolvents slightly beyond it,
it would indicate the non-fluctuating value $0$. This is 
because the wave function of her resolvent would no longer be orthogonal
to that of the  ground state.
She would conclude that there is something classical, a wall. 
It resembles the heuristic example of a true condensate
on a wall. 

\mn 
\bA  I am skeptical. This model does not describe interaction. 

\section{Epilogue}
\label{sec3}
\setcounter{equation}{0} 

This transcript reflects the views of Alice and Bob. It is,
of course, a condensate of many discussions. It seems as though there is
no prospect of reaching a consensus between them. 
Alice will probably
continue to use the OP-criterion in her work since it fits
with her intuitive picture of a true condensate.
Bob will continue to regard the PC condition as conceptually superior to
the OP criterion, without
questioning its computational aspects.

\medskip
It is noteworthy that the mathematical framework, which led to this
controversy, the resolvent algebra \cite{BuGr}, was never an issue
in these discussions.
This algebra is the answer to the age-old question of whether the
observables in quantum mechanics can be described by the elements
of a universal algebra on which the many dynamics of interest act. In essence,
this was Heisenberg's original idea. Yet, some time after the invention of
quantum mechanics, the 
search for such an algebra was given up. Polynomials of position and
momentum operators did not work and their exponentials, the
Weyl operators, did not work either.

\medskip
So one resorted to a dual
picture, the Hilbert spaces of states. Assuming that observables do
not change particle numbers (though they do in high energy physics), 
one considered $n$-particle spaces on which the position and
momentum operators act by unbounded operators. And, happily, for
given $n$ these representations are unique according to the
Stone-von Neumann theorem. So the dynamics could be described by
Hamiltonians, which depend on $n$. 

\medskip 
But the problem reappears if one proceeds to infinite $n$, as for
example in the OP-criterion. Then Stone and von Neumann no longer help.
There appear zillions of states which differ by
classical observables. Well known examples 
are temperature distributions and the spatial structure of
phases. It remained 
unclear whether there is some quantum algebra which
describes for given dynamics these classical observables in the
limit of large $n$.

\medskip
After the discovery of particle statistics, it became clear that it is
convenient to use a field-theoretic description of the particles, at least
as a bookkeeping device. So for bosons, one considers Bose fields that
satisfy canonical commutation relations.
Observables, which do not change the particle numbers, are combinations
of the fields which commute with the particle number operator. This
insight enters for example in the OP-criterion. But the question
of whether there
is a version of this algebra that is stable under the dynamics remained.

\medskip
Such an algebra, the resolvent algebra, was presented more than 80
years after Heisenberg's groundbreaking insights \cite{BuGr}.
The algebra of observables, which covers all finite and infinite
particle systems, was subsequently introduced in \cite{Bu1} using Bose fields.
It is stable
under the action of the dynamics induced by a large family of
two-body interactions, which includes attractive and repulsive potentials 
and potentials of short and long range. But there is another
important structural feature 
of this  algebra of observables. It is the fact that it contains ideals, \ie
proper subalgebras which are stable under left and right multiplication
by all elements of the full algebra.

\medskip 
It was clear from
the outset that an algebra which is stable under the action of
a variety of dynamics must have ideals \cite{BuGr}. But these ideals play
also an important role in the interpretation of the limit states.
They are the ingredients which describe classical features
in the limit of large particle numbers.
It has been shown in \cite{Bu6} that any basic resolvent
with a given wave function determines such an ideal. Mathematics
tells us that on any algebra there exist states where all elements of an 
ideal have the expectation
value $0$, as for proper condensates in the PC-condition.
So the possible appearance of such condensates is 
already encoded in the algebraic structures. 

\medskip
The resolvent algebra provides a complementary view on many-body systems.
It would make sense to refer to it as
\textit{algebraic many-body theory}, in
reference to \mbox{\textit{algebraic quantum field theory}}. The latter approach
has produced quite a few deep  results. It furthered our understanding
of relativistic many-body systems, described by quantum fields. These
results were not derived by the study of a family of fully consistent models,
which did not exist. They were based on general physical principles,
such as the condition of locality for observables, which is the algebraic
version of the principle of Einstein causality \cite{Ha}.

\medskip
In case of the resolvent algebra, one is in a much better 
situation. One does not need to rely on general principles because the
resolvent algebra is concretely given and accommodates a large family
of different dynamics. Nevertheless, I think that, 
rather than restricting oneself to analyzing individual models,
one should adopt a more global perspective.

\medskip
Let me indicate by an example of what I have in mind. Recently, it was shown
in \cite{Bu6} that canonical ensembles of $n$ particles, which are
confined by a fixed
harmonic potential and interact with each other, approach in the limit
of large $n$ states which are still in equilibrium. They satisfy the
so-called KMS-condition \cite{Ha}, which characterizes equilibria
of infinite systems. This result did not follow from
a detailed study of the large family of admitted dynamics. It is the
consequence of general results in the theory of operator algebras
\cite{BrRoII}.

\medskip
Knowing that the ensembles have physically meaningful limits, the next
question is: do there appear condensates? Heuristically,
one may expect that the sequence of states satisfies 
the PC-condition quite generally. In case of short range
attractive or repulsive forces, the particles should form
increasing clusters around the bottom of the harmonic
potential which become heavy classical bodies at rest in the limit. 
Their impact on the remaining particles, if any, would be an
additional external force.

\medskip 
On the theoretical side, scaled resolvents with wave functions,
having support about
the bottom, should yield expectation values which are less than  
$1$ in the limit; this would indicate an OP-condensate.
Alice would not be surprised by this result
since the harmonic potential is kept fixed. 
On the other hand, in view of the new additional external force,
forming streams of particles, there could be (unscaled) resolvents
with wave functions which do not converge to $0$ in the limit. 
This would please Bob since it
would show that the sequence describes a proper condensate.

\medskip
Most likely, Alice will not venture into this field, and Bob
has already left~it. He is now working on another conceptual problem.
Age also plays a role. As for me, I am confident that
algebraic many-body theory will attract younger mathematical physicists,
who wish to contribute to the conceptual and mathematical
foundations of many-body physics. Time will tell.

\bigskip 
\noindent {\Large \bf Acknowledgment} \\[1mm]
I would like to thank Jakob Yngvason for our many discussions on this
topic and his encouragement to write this fictional transcript.

\end{document}